\documentclass[12pt]{article}
\usepackage{bbm,latexsym}
\usepackage{amssymb,amsmath}
\usepackage{xcolor} 
\usepackage{slashed}
\newcommand{\ve}{\varepsilon}
\newcommand{\coe}{\mathcal{O}(\varepsilon)}
\newcommand{\con}{\mathcal{O}(\eta)}
\newcommand{\dd}{\mathrm{d}}
\newcommand{\bone}{\mathbbm{1}}
\newcommand{\diag}{\mathrm{diag}}
\newcommand{\zz}{Z^{(1/2)}}
\newcommand{\cpt}{(\mathcal{CPT})}
\allowdisplaybreaks
\begin{document}

\title{
\normalsize \hfill UWThPh-2016-11 \\[10mm]
\LARGE Revisiting on-shell renormalization conditions \\
in theories with flavour mixing}

\author{
W.~Grimus\thanks{E-mail: walter.grimus@univie.ac.at}\;
\addtocounter{footnote}{1}
and M.~L\"oschner\thanks{E-mail: maximilian.loeschner@univie.ac.at}
\\[5mm]
\small University of Vienna, Faculty of Physics \\
\small Boltzmanngasse 5, A--1090 Vienna, Austria
}

\date{June 20, 2016}

\maketitle

\begin{abstract}
In this review, 
we present a derivation 
of the on-shell renormalization conditions for scalar 
and fermionic fields in theories with and without parity conservation. 
We also discuss the specifics of Majorana fermions.
Our
approach only assumes a canonical form for the renormalized propagators and
exploits the fact that the inverse propagators 
are non-singular in $\ve = p^2 - m_n^2$, where $p$ is the external
four-momentum and $m_n$ is a pole mass. 
In this way, 
we obtain full agreement
with commonly used on-shell conditions.
We also discuss how they are implemented in renormalization.
\end{abstract}

\newpage

\section{Introduction}
On-shell renormalization conditions in theories with inter-family or flavour
mixing~\cite{donoghue,aoki,soares,denner,kniehl1996,kniehl2012} 
are quite important in view of the experimentally established quark and lepton
mixing matrices~\cite{rpp} and the mixing between the photon and the $Z$ boson
in the Standard Model---see for instance~\cite{sirlin,hollik}. In theories
beyond the Standard Model, mixing of new 
fermions and scalar mixing might occur as well.
However, we feel that several aspects of the derivation of the on-shell
renormalization conditions remain a bit vague for the general reader of the
relevant literature and should be discussed in more detail.
In this review, we present a consistent way of deriving the on-shell
conditions, first for scalar, then for fermionic fields in theories with and
without parity conservation and finally for Majorana fields,
for the mixing of $N$ fields. Our approach is solely based
on the pole structure of the $N \times N$ propagator matrix and relies on the
fact that the inverse propagator has no singularity in $p^2$, where $p$ is the
external momentum. We also review the
counting of the number of renormalization conditions and demonstrate that this
number coincides with the number of degrees of freedom in the
counterterms, except for overall phase factors that remain free for the
fermionic fields in the case of Dirac fermions. 
Our results agree, of course, with the ones derived
in~\cite{aoki}. Explicit formulas for the field strength renormalization
constants and mass counterterms in the above-mentioned theories are given for
the lowest non-trivial order. 

Our discussion is based on two assumptions:
\begin{enumerate}
\item
All $N$ physical masses $m_i$ are different.
\item
The $N$ poles in the propagator matrix are located in a region where
absorptive parts are absent or can be neglected.
\end{enumerate}
Some remarks relating to these assumptions are in order. 
Equality of two or more masses would require a ``flavour'' symmetry, but
dealing with on-shell renormalization in the presence of such a symmetry is
beyond the scope of this review.\footnote{This would, for instance, change the
  aforementioned counting of number of renormalization conditions and degrees
  of freedom in the counterterms.} 
The on-shell renormalization in this review deals with the dispersive
parts of the scalar and fermionic propagators. 
If large imaginary parts appear in the higher order corrections to the
propagators, a treatment of their absorptive parts becomes relevant, but lies
beyond the scope of this review as well. 
Still, the renormalization conditions derived here are fully applicable in the
regions of $p^2$ where the absorptive parts vanish. 
Elsewhere, they can be used by simply inserting only the dispersive parts of the
self-energy functions into the conditions. 
Another possible approach is to use complex masses as well as complex
counterterms in the so-called complex-mass scheme. 
A treatment of this approach can \textit{e.g.}\ be found
in~\cite{Denner:2006ic,Denner:1999gp}. Since our paper is intended as a
pedagogical review and the distinction between dispersive an absorptive
parts in the propagator matrix plays an important role in our presentation, we
have included, for the sake of completeness, a discussion of this issue as an
appendix. 

The treatment of on-shell conditions in our paper is based on an expansion in 
$p^2 - m_n^2$ around each pole mass $m_n$, for both propagator and
inverse propagator, whereas the authors of~\cite{kniehl2012} make use of 
exact matrix relations between the propagator matrix and its inverse. In this
sense, our treatment of fermions is complementary to that
of~\cite{kniehl2012}. 

The plan of the paper is as follows. In section~\ref{scalar} we discuss mixing
of real scalar fields, whereas fermions are treated in section~\ref{fermion}
in the case of parity conservation. The complications which arise
when parity is violated are elaborated in 
section~\ref{fermion-parity violation}. This section contains also a
subsection on the on-shell renormalization of Majorana fermions.
After a summary in section~\ref{summary}, the emergence of dispersive and
absorptive contributions to the propagator is covered in
appendix~\ref{dispersive} in the framework of the K\"all\'en--Lehmann
representation. Some computational details in the treatment of fermions are
deferred to appendix~\ref{computational}. The condition on the propagator
matrix which arises in the case of Majorana nature of the fermions is derived
in appendix~\ref{Majorana condition}.

In the following, we will use $k,l$ as summation indices whereas $i,j,n$ do
not imply summation.

\section{Scalar propagator}
\label{scalar}
\subsection{On-shell conditions}
We first study the scalar propagator for $N$ real scalar
fields, which is a simple and instructive case to begin with. 
Here the propagator is an $N \times N$ matrix
\begin{equation}
\Delta(p^2) = \left( \Delta_{ij}(p^2) \right),
\end{equation}
where $p^2$ is the Minkowski square of the four-momentum $p$. We assume that
all masses $m_n$ of the scalars are different. 
We stress that $\Delta(p^2)$ is the \emph{renormalized} propagator.
Defining
\begin{equation}\label{eps}
\ve \equiv p^2 - m_n^2,
\end{equation}
the on-shell renormalization conditions consist of the requirement~\cite{aoki} 
\begin{equation}\label{onshell}
\Delta_{ij}(p^2) \stackrel{\ve \to 0}{\longrightarrow} 
\frac{\delta_{in} \delta_{nj}}{\ve} + \Delta^{(0)}_{ij} + \coe
\end{equation}
for all $n = 1, \ldots, N$. In this formula and in the following,
the symbol $\delta_{rs}$ always signifies the Kronecker symbol. 
The coefficients $\Delta^{(0)}_{ij}$ are of order one \textit{viz.}\ $\ve^0$. 
In the following, the 
superscripts $(0)$ and $(1)$ will always indicate order $\ve^0$ and $\ve^1$,
respectively. 

The complication comes from the fact that we actually want to impose
on-shell renormalization conditions on the \emph{inverse} propagator, which
we denote by $A = \left( A_{ij} \right)$.\footnote{From now on, for the sake of
  simplicity of notation, we skip the dependence on $p^2$ in all quantities,
  whenever this dependence is obvious.}
Accordingly, we have to translate equation~(\ref{onshell}) into conditions on
$A$. The inverse propagator fulfills 
\begin{equation}\label{da}
\Delta_{ik} A_{kj} = A_{ik} \Delta_{kj} = \delta_{ij} 
\quad \mbox{with} \quad 
A_{ij} =  A^{(0)}_{ij} + \ve A^{(1)}_{ij} + \mathcal{O}(\ve^2),  
\end{equation}
where the latter relation states that $A$ has no singularity in $\ve$.
Equation~(\ref{da}) is reformulated as
\begin{subequations}\label{o}
\begin{eqnarray}
&&
\Delta_{ik} A_{kj} = 
\frac{1}{\ve} \delta_{in} A^{(0)}_{nj} + 
\delta_{in} A^{(1)}_{nj} + \Delta^{(0)}_{ik} A^{(0)}_{kj} + \coe =
\delta_{ij},
\\
&&
A_{ik} \Delta_{kj} = 
\frac{1}{\ve} A^{(0)}_{in} \delta_{nj} + 
A^{(1)}_{in} \delta_{nj} + A^{(0)}_{ik} \Delta^{(0)}_{kj} + \coe =
\delta_{ij}.
\end{eqnarray}
\end{subequations}
Avoiding the singularity in $1/\ve$ requires
\begin{equation}\label{sc1}
A^{(0)}_{in} = 0 \;\; \forall \; i=1,\ldots,N
\quad \mbox{and} \quad
A^{(0)}_{nj} = 0 \;\; \forall \; j=1,\ldots,N.
\end{equation}
For $i=j=n$ we obtain the further condition
\begin{equation}\label{sc2}
A^{(1)}_{nn} = 1.
\end{equation}
Note that, because of equation~(\ref{sc1}), 
$\Delta^{(0)}_{nk}$ and $\Delta^{(0)}_{kn}$ do not occur in the second condition.

The remaining coefficients in equation~(\ref{o}), which have not yet
been fixed by equations~(\ref{sc1}) and~(\ref{sc2}), are determined by the
orthogonality conditions at order $\ve^0$:
\begin{equation}
\renewcommand{\arraystretch}{1.5}
\begin{array}{rl}
i \neq n, \, j \neq n: &
\Delta^{(0)}_{ik} A^{(0)}_{kj} = A^{(0)}_{ik} \Delta^{(0)}_{kj} = \delta_{ij},
\\
i = n, \, j \neq n: &
A^{(1)}_{nj} + \Delta^{(0)}_{nk} A^{(0)}_{kj} = 0,
\\
i \neq n, \, j = n: &
A^{(1)}_{in} + A^{(0)}_{ik} \Delta^{(0)}_{kn} = 0.
\end{array}
\end{equation}
However, these conditions have nothing to do with on-shell renormalization.

In summary, for the inverse propagator we have derived the on-shell conditions 
\begin{equation}
A_{in}(m_n^2) = A_{nj}(m_n^2) = 0 \;\; \forall\; i,j=1,\ldots,N
\quad \mbox{and} \quad
\left. \frac{\dd A_{nn}(p^2)}{\dd p^2} \right|_{p^2 = m_n^2} = 1,
\end{equation}
in agreement with the result of~\cite{aoki}.
Of course, for every $n$ there is such a set of conditions.

\subsection{Renormalization and parameter counting}
\label{scalar-parameters}

The dispersive part of the inverse propagator is a real
symmetric matrix, \textit{i.e.}
\begin{equation}\label{eq:scalar-prop-symm}
\left( \Delta^{-1}(p^2) \right)^T = \Delta^{-1}(p^2).
\end{equation}
A derivation of this symmetry relation in terms of
the K\"all\'en--Lehmann representation of the scalar propagator is given in
appendix~\ref{sec:scalar-prop-symmetry}. 
Using the fact that on-shell renormalization conditions are imposed on 
the dispersive part, we can reformulate them as
\begin{equation}\label{cond-scalar}
i \neq j\!: \; A_{ij}(m_j^2) = 0, \quad
i = j\!: \; A_{ii}(m_i^2) = 0, \; 
\left. \frac{\dd A_{ii}(p^2)}{\dd p^2} \right|_{p^2 = m_i^2} = 1.
\end{equation}
The number of on-shell conditions is thus
\begin{equation}
2\, \binom{N}{2} + 2 N = N^2 + N.
\end{equation}
The factor 2 in front of the binomial coefficient originates from the two pairs
$(i,j)$ and $(j,i)$ for $i \neq j$.

In terms of the self-energy, the renormalized inverse propagator has the form
\begin{equation}
A(p^2) = p^2 - {\hat m}^2 - \Sigma^{(r)}(p^2).
\end{equation}
The bare fields $\varphi^{(b)}_i$ are related to the renormalized ones
via the field strength renormalization matrix:
\begin{equation}
\varphi^{(b)}_i = \left( \zz \right)_{ik} \varphi_k.
\end{equation}
In this way, the renormalized self-energy $\Sigma^{(r)}(p^2)$ is related to
the unrenormalized one by
\begin{equation}
\Sigma^{(r)}(p^2) = \Sigma(p^2) + 
\left( \bone - \left( \zz \right)^T Z^{(1/2)} \right) p^2 + 
\left( \zz \right)^T \left( {\hat m}^2 + \delta {\hat m}^2
\right) \zz - {\hat m}^2.
\end{equation}
For real scalar fields, $\zz$ is a general real $N \times N$
matrix. Therefore, it contains $N^2$ real parameters. 
Assuming to be in a basis where
\begin{equation}
{\hat m}^2 = \diag \left( m_1^2, \ldots, m_N^2 \right),
\end{equation}
then $\delta {\hat m}^2$ is diagonal too.
Thus, there are
$N$ real parameters in $\delta {\hat m}^2$. In summary, we have $N^2 + N$ 
free parameters at disposal for implementing $N^2 + N$ on-shell renormalization
conditions. The condition 
$A_{ii}(m_i^2) = 0$ 
is imposed with the help of 
$\delta {\hat m}^2_i$, while for the rest the field strength renormalization
matrix $\zz$ is responsible.

It is instructive to perform the renormalization of the inverse propagator at
the lowest non-trivial order in $\zz$. In this case we write 
\begin{equation}
\zz = 1 + \frac{1}{2}\,z
\end{equation}
and
\begin{equation}
\Sigma^{(r)}(p^2) = \Sigma(p^2) - 
\frac{1}{2} \left( z + z^T \right) p^2 + 
\frac{1}{2} \left( {\hat m}^2 z + z^T {\hat m}^2 \right) + \delta {\hat m}^2.
\end{equation}
Then
it is straightforward to derive 
\begin{equation}
i \neq j\!: \; \frac{1}{2}\, z_{ij} = 
-\frac{\Sigma_{ij}(m_j^2)}{m_i^2 - m_j^2}, \quad
i = j\!: \; \delta {\hat m}^2_i = -\Sigma_{ii}(m_i^2), \;
z_{ii} = \left. \frac{\dd \Sigma_{ii}(p^2)}{\dd p^2} \right|_{p^2 =  m_i^2}
\end{equation}
from equation~(\ref{cond-scalar}).

\section{Fermion propagator with parity conservation}
\label{fermion}
\subsection{On-shell conditions}
For fermions the situation is more involved, but we can nevertheless proceed
in analogy to the scalar case. We write the propagator as
\begin{equation}
S = C \slashed{p} - D
\quad \mbox{with} \quad
C = \left( C_{ij}(p^2) \right), \quad D = \left( D_{ij}(p^2) \right)
\end{equation}
being $N \times N$ matrices. It is now expedient to formulate for fermions the
condition analogous to equation~(\ref{onshell}).
Using again $\ve$ of equation~(\ref{eps}), we have now
\begin{equation}\label{f-onshell}
S_{ij} \stackrel{\ve \to 0}{\longrightarrow} 
\frac{\delta_{in} \delta_{nj}}{\slashed{p} - m_n} + \tilde S_{ij},
\end{equation}
where $\tilde S_{ij}$ is non-singular in $\ve$.
First of all, we have to work out what equation~(\ref{f-onshell}) means for $C$
and $D$. Multiplying $S$ by $\slashed{p} - m_n$ and exploiting
equation~(\ref{f-onshell}), we find 
\begin{equation}
\left( \slashed{p} - m_n \right) S_{ij} = 
\delta_{in} \delta_{nj} + \left( \slashed{p} - m_n \right) \tilde S_{ij} =
\ve C_{ij} - 
\left( \slashed{p} - m_n \right) \left( D_{ij} + m_n C_{ij} \right),
\end{equation}
whence we conclude
\begin{equation}\label{cd}
C_{ij} = \frac{\delta_{in} \delta_{nj}}{\ve} + C^{(0)}_{ij} + \coe,
\quad
D_{ij} = -\frac{m_n \delta_{in} \delta_{nj}}{\ve} + D^{(0)}_{ij} + \coe.
\end{equation}
The second relation follows from the non-singularity of $\tilde S_{ij}$.
We see that in the fermion case we have two relations instead of one,
equation~(\ref{onshell}), in the scalar case.

Now we have to formulate the conditions of equation~(\ref{cd}) for the inverse
propagator 
\begin{equation}
S^{-1} = A \slashed{p} - B.
\end{equation}
The relation between $A$, $B$ and $C$, $D$ is given by
\begin{subequations}\label{ss}
\begin{eqnarray}
\left( S S^{-1} \right)_{ij} = \delta_{ij}
& \Rightarrow & 
C_{ik} A_{kj}\, p^2 + D_{ik} B_{kj}  = \delta_{ij}, \quad
C_{ik} B_{kj}\, + D_{ik} A_{kj}  = 0,
\\
\left( S^{-1} S \right)_{ij} = \delta_{ij}
& \Rightarrow & 
A_{ik} C_{kj}\, p^2 + B_{ik} D_{kj}  = \delta_{ij}, \quad
B_{ik} C_{kj}\, + A_{ik} D_{kj}  = 0.
\end{eqnarray}
\end{subequations}
Since the inverse propagator is non-singular for $\ve \to 0$,
we have the expansion
\begin{equation}\label{ab}
A_{ij} =  A^{(0)}_{ij} + \ve A^{(1)}_{ij} + \mathcal{O}(\ve^2),
\quad
B_{ij} =  B^{(0)}_{ij} + \ve B^{(1)}_{ij} + \mathcal{O}(\ve^2)
\end{equation}
with 
\begin{equation}
A^{(0)}_{ij} = A_{ij}(m_n^2),
\quad 
A^{(1)}_{ij} = \left. \frac{\dd A_{ij}(p^2)}{\dd p^2} \right|_{p^2 = m_n^2} 
\end{equation}
and the analogous relations for $B$.
We have to plug the relations of equation~(\ref{ab}) into equation~(\ref{ss})
and invoke the expansion for the propagator presented in equation~(\ref{cd}). 
For the details, we refer the reader to appendix~\ref{app-pc}.

Summarizing the computation in appendix~\ref{app-pc}, the on-shell
renormalization conditions on the renormalized inverse propagator are given by
\begin{subequations}\label{fc}
\begin{eqnarray}
&& \label{in}
B_{in}(m_n^2) =  m_n A_{in}(m_n^2) \;\; \forall\; i = 1,\ldots,N;
\\
&& \label{nj}
B_{nj}(m_n^2) =  m_n A_{nj}(m_n^2) \;\; \forall\; j = 1,\ldots,N;
\\ 
&& \label{ann}
A_{nn}(m_n^2) + 
2m_n^2 \left. \frac{\dd A_{nn}(p^2)}{\dd p^2} \right|_{p^2 = m_n^2} - 
2m_n \left. \frac{\dd B_{nn}(p^2)}{\dd p^2} \right|_{p^2 = m_n^2} = 1.
\end{eqnarray}
\end{subequations}
They are in agreement with the conditions derived in~\cite{aoki}.
Relations~(\ref{in}) and~(\ref{nj}) follow from the cancellation of the terms
with $1/\ve$, whereas relation~(\ref{ann}) stems from the terms with 
zeroth power in $\ve$.

\subsection{Renormalization and parameter counting}
\label{rp-parityconservation}
As in the scalar case, we can make use of a symmetry of the dispersive part of
the propagator, namely
\begin{equation}\label{Scond}
\gamma_0 \left( S^{-1}(p) \right)^\dagger_\mathrm{disp} \gamma_0 = 
S^{-1}(p)_\mathrm{disp},
\end{equation}
which shows us that
\begin{equation}\label{AB}
A^\dagger = A, \quad B^\dagger = B
\end{equation}
is valid for the dispersive parts. 
A derivation of equation~(\ref{Scond}) is given
in appendix~\ref{sec:fermion-prop-symmetry}. 
Therefore, the on-shell conditions can be rewritten as 
\begin{subequations}\label{ABij}
\begin{eqnarray}
\label{ijAB}
i \neq j\!: && B_{ij}(m_j^2) = m_j A_{ij}(m_j^2), 
\\
\label{i=j}
i = j\!: && B_{ii}(m_i^2) = m_i A_{ii}(m_i^2), 
\\
\label{i=j-d}
&&
A_{ii}(m_i^2) + 
2 m_i^2 \left. \frac{\dd A_{ii}(p^2)}{\dd p^2} \right|_{p^2 =m_i^2} - 
2 m_i \left. \frac{\dd B_{ii}(p^2)}{\dd p^2} \right|_{p^2 = m_i^2} = 1.
\end{eqnarray}
\end{subequations}
For each pair $i \neq j$ we have four conditions, because
relation~(\ref{ijAB}) is complex and contains $i < j$ and $i > j$.
For every $i=j$ there are two real conditions, one from equation~(\ref{i=j})
and one from equation~(\ref{i=j-d}). Thus, the number of on-shell
conditions amounts to 
\begin{equation}\label{number}
4\, \binom{N}{2} + 2N = 2N^2.
\end{equation}

In analogy to the scalar case, we write
\begin{equation}\label{invSF}
S^{-1}(p) = \slashed{p} - {\hat m} - \Sigma^{(r)}(p).
\end{equation}
We introduce the field strength renormalization via
\begin{equation}
\psi^{(b)}_i = \left( \zz \right)_{ik} \psi_k
\end{equation}
with bare and renormalized fields $\psi^{(b)}_i$ and $\psi_k$,
respectively. Then, the renormalized self-energy $\Sigma^{(r)}(p)$
is written as 
\begin{equation}
\Sigma^{(r)}(p) = \Sigma(p) +  
\left( \bone - \left( \zz \right)^\dagger Z^{(1/2)} \right) \slashed{p} + 
\left( \zz \right)^\dagger \left( \hat m + \delta \hat m \right) \zz - \hat m
\end{equation}
with the unrenormalized self-energy $\Sigma(p)$.
From this equation it is obvious that the transformation
\begin{equation}\label{phasefactors}
\zz \to e^{i\hat \alpha} \zz
\end{equation}
by a diagonal matrix $e^{i\hat \alpha}$ of phase factors leaves  
$\Sigma^{(r)}(p)$ invariant. Though $\zz$ is a general complex $N \times N$
matrix, we only have $2N^2 -N$ parameters in this matrix at our
disposal. Adding to this number the $N$ parameters in $\delta \hat m$, we
arrive at $2 N^2$ renormalization parameters, which equals
the number of renormalization conditions in equation~(\ref{number}).

To perform the renormalization, we decompose the unrenormalized self-energy 
into
\begin{equation}
\Sigma(p) = \Sigma^{(A)}(p^2) \slashed{p} + \Sigma^{(B)}(p^2).
\end{equation}
This leads to 
\begin{equation}
A = \left( \zz \right)^\dagger Z^{(1/2)} - \Sigma^{(A)} 
\quad \mbox{and} \quad
B = \Sigma^{(B)} +
\left( \zz \right)^\dagger \left( \hat m + \delta \hat m \right) \zz.
\end{equation}

Let us---just as in the scalar case---do the one-loop renormalization to
illustrate the above discussion.
Again we set $\zz = \bone + \frac{1}{2}\,z$ and obtain
\begin{subequations}
\begin{eqnarray}
A &=& \bone - \Sigma^{(A)} + \frac{1}{2} \left( z + z^\dagger \right),
\\
B &=& \hat m + \Sigma^{(B)} + 
\frac{1}{2} \left( \hat m z + z^\dagger \hat m \right) + 
\delta \hat m.
\end{eqnarray}
\end{subequations}
We assume a diagonal mass matrix
\begin{equation}
\hat m = \diag \left( m_1, \ldots, m_N \right).
\end{equation}
Inserting these quantities $A$, $B$ and $\delta \hat m$ into
equation~(\ref{ABij}), it is straightforward to find the solution
\begin{equation}
\frac{1}{2} z_{ij} = -\frac{1}{m_i - m_j} \left( 
m_j \Sigma^{(A)}_{ij}(m_j^2) + \Sigma^{(B)}_{ij}(m_j^2) \right)
\end{equation}
for $i \neq j$ and
\begin{subequations}
\begin{eqnarray}
\delta m_i &=& - m_i \Sigma^{(A)}_{ii}(m_i^2) - \Sigma^{(B)}_{ii}(m_i^2),
\\
\mathrm{Re}\,z_{ii} &=& 
\Sigma^{(A)}_{ii}(m_i^2) + 
2 m_i^2 \left. \frac{\dd \Sigma^{(A)}_{ii}(p^2)}{\dd p^2} \right|_{p^2 = m_i^2} + 
2 m_i \left. \frac{\dd \Sigma^{(B)}_{ii}(p^2)}{\dd p^2} \right|_{p^2 = m_i^2}
\end{eqnarray}
\end{subequations}
for $i=j$.
We notice that $\mathrm{Im}\,z_{ii}$ is not determined as a consequence of the
phase freedom expressed in equation~(\ref{phasefactors}).

\subsection{Formal derivation of the fermionic on-shell conditions}
An interesting aspect of the on-shell conditions for fermions is that they can
be derived by formally considering $\slashed{p}$ as a variable~\cite{aoki} and
expanding in 
\begin{equation}
\eta \equiv \slashed{p} - m_n.
\end{equation}
Then one can exactly imitate the computation for scalars.
We begin with the condition 
\begin{equation}
S_{ij} \stackrel{\eta \to 0}{\longrightarrow} 
\frac{\delta_{in} \delta_{nj}}{\eta} + S^{(0)}_{ij}
\end{equation}
for the propagator. Then with 
\begin{equation}
T = S^{-1} \quad \mbox{and} \quad
T_{ij} = T^{(0)}_{ij} + \eta T^{(1)}_{ij} + \mathcal{O}(\eta^2)
\end{equation}
we obtain
\begin{subequations}
\begin{eqnarray}
&&
S_{ik} T_{kj} = 
\frac{1}{\eta} \delta_{in} T^{(0)}_{nj} + 
\delta_{in} T^{(1)}_{nj} + S^{(0)}_{ik} T^{(0)}_{kj} + \con =
\delta_{ij},
\\
&&
T_{ik} S_{kj} = 
\frac{1}{\eta} T^{(0)}_{in} \delta_{nj} + 
T^{(1)}_{in} \delta_{nj} + T^{(0)}_{ik} S^{(0)}_{kj} + \con =
\delta_{ij}.
\end{eqnarray}
\end{subequations}
In this way we arrive at conditions completely analogous to the scalar
conditions: 
\begin{equation}\label{t}
T^{(0)}_{in} = 0 \;\; \forall i=1,\ldots,N, \quad
T^{(0)}_{nj} = 0 \;\; \forall j=1,\ldots,N, \quad
T^{(1)}_{nn} = 1.
\end{equation}
But we know that the inverse propagator has the decomposition
\begin{equation}
T_{ij}(\slashed{p}) = A_{ij}(p^2) \slashed{p} - B_{ij}(p^2),
\end{equation}
where $p^2 = \slashed{p}^2$.
\begin{subequations}
Therefore, equation~(\ref{t}) translates into 
\begin{eqnarray}
&&
T_{in} (\slashed{p}=m_n) = A_{in}(m_n^2) m_n - B_{in}(m_n^2) = 0,
\\
&&
T_{nj} (\slashed{p}=m_n) = A_{nj}(m_n^2) m_n - B_{nj}(m_n^2) = 0.
\end{eqnarray}
\end{subequations}
In order to tackle $T^{(1)}_{nn} = 1$, we note that
\begin{equation}
\frac{\dd p^2}{\dd \slashed{p}} = 2\slashed{p}.
\end{equation}
Therefore, we end up with
\begin{equation}
\left. \frac{\dd T_{nn}(\slashed{p})}{\dd \slashed{p}}
\right|_{\slashed{p}=m_n} = A_{nn}(m_n^2) + 
2m_n^2 \left. \frac{\dd A_{nn}(p^2)}{\dd p^2} \right|_{p^2 = m_n^2} - 
2m_n \left. \frac{\dd B_{nn}(p^2)}{\dd p^2} \right|_{p^2 = m_n^2} = 1.
\end{equation}
Thus we recover relation~(\ref{ann}).

\section{Fermion propagator without parity conservation}
\label{fermion-parity violation}
\subsection{On-shell conditions}
With the chiral projectors
\begin{equation}
\gamma_L = \frac{\bone - \gamma_5}{2}, \quad 
\gamma_R = \frac{\bone + \gamma_5}{2},
\end{equation}
the propagator is now given by
\begin{equation}
S = \slashed{p} \left( C_L \gamma_L + C_R \gamma_R \right) - 
\left( D_L \gamma_L + D_R \gamma_R \right).
\end{equation}
The on-shell condition~(\ref{f-onshell}) is again valid.
Since
\begin{equation}
\left( \slashed{p} - m_n \right) S_{ij} = 
\ve \left( C_L \gamma_L + C_R \gamma_R \right)_{ij} - 
\left( \slashed{p} - m_n \right) 
\left[ \left( D_L + m_n C_L \right) \gamma_L + 
\left( D_R + m_n C_R \right) \gamma_R \right]_{ij},
\end{equation}
we have the following behaviour for $\ve \to 0$:
\begin{subequations}\label{prop-exp}
\begin{eqnarray}
(C_L)_{ij} = \frac{\delta_{in} \delta_{nj}}{\ve} + 
(C^{(0)}_L)_{ij} + \coe,
&&
(D_L)_{ij} = -\frac{m_n \delta_{in} \delta_{nj}}{\ve} + 
(D^{(0)}_L)_{ij} + \coe,
\\
(C_R)_{ij} = \frac{\delta_{in} \delta_{nj}}{\ve} + 
(C^{(0)}_R)_{ij} + \coe,
&&
(D_R)_{ij} = -\frac{m_n \delta_{in} \delta_{nj}}{\ve} + 
(D^{(0)}_R)_{ij} + \coe.
\end{eqnarray}
\end{subequations}

We define the inverse propagator as
\begin{equation}\label{Sinv}
S^{-1} = \slashed{p} \left( A_L \gamma_L + A_R \gamma_R \right) - 
\left( B_L \gamma_L + B_R \gamma_R \right).
\end{equation}
Since the set of matrices 
\begin{equation}
\left\{ \gamma_L,\,\gamma_R,\, 
\slashed{p} \gamma_L,\,\slashed{p} \gamma_R \right\}
\end{equation}
is linearly independent for $p \neq 0$,
we obtain the relations
\begin{subequations} \label{eq:prop-times-inverseprop}
\begin{eqnarray}
\left( S S^{-1} \right)_{ij} = \delta_{ij}
& \Rightarrow & 
\left( C_R A_L\, p^2 + D_L B_L \right)_{ij} = \delta_{ij}, \quad
\left( C_L B_L + D_R A_L \right)_{ij}  = 0,
\\
&&
\left( C_L A_R\, p^2 + D_R B_R \right)_{ij} = \delta_{ij}, \quad
\left( C_R B_R + D_L A_R \right)_{ij}  = 0,
\\
\left( S^{-1} S \right)_{ij} = \delta_{ij}
& \Rightarrow & 
\left( A_R C_L\, p^2 + B_L D_L \right)_{ij}  = \delta_{ij}, \quad
\left( B_R C_L + A_L D_L \right)_{ij}  = 0,
\\
&&
\left( A_L C_R\, p^2 + B_R D_R \right)_{ij}  = \delta_{ij}, \quad
\left( B_L C_R + A_R D_R \right)_{ij}  = 0.
\end{eqnarray}
\end{subequations}
Because the inverse propagator is non-singular in $\ve$, we are allowed to
assume the expansion
\begin{subequations}\label{invprop-exp}
\begin{eqnarray}
(A_L)_{ij} =  (A^{(0)}_L)_{ij} + \ve (A^{(1)}_L)_{ij} + \mathcal{O}(\ve^2),
&&
(B_L)_{ij} =  (B^{(0)}_L)_{ij} + \ve (B^{(1)}_L)_{ij} + \mathcal{O}(\ve^2),
\hphantom{xx}\\
(A_R)_{ij} =  (A^{(0)}_R)_{ij} + \ve (A^{(1)}_R)_{ij} + \mathcal{O}(\ve^2),
&&
(B_R)_{ij} =  (B^{(0)}_R)_{ij} + \ve (B^{(1)}_R)_{ij} + \mathcal{O}(\ve^2).
\end{eqnarray}
\end{subequations}
For the details of evaluating equations~(\ref{prop-exp}),
(\ref{eq:prop-times-inverseprop}) and~(\ref{invprop-exp}), we refer the reader
to appendix~\ref{app-without}.

In summary, the on-shell conditions on the inverse propagator are given by
equations~(\ref{abab}) and~(\ref{=1}). 
The first equation reads
\begin{subequations}\label{final-c1}
\begin{eqnarray}
(B_L)_{in}(m_n^2) =  m_n (A_R)_{in}(m_n^2), \quad 
(B_R)_{in}(m_n^2) =  m_n (A_L)_{in}(m_n^2) 
&& \forall\; i, \label{final-c1-a}
\\
(B_L)_{nj}(m_n^2) =  m_n (A_L)_{nj}(m_n^2), \quad
(B_R)_{nj}(m_n^2) =  m_n (A_R)_{nj}(m_n^2) 
&& \forall\; j. \label{final-c1-b}
\end{eqnarray}
\end{subequations}
Equation~(\ref{=1}) can be brought into the form
\begin{subequations}\label{eq:residue-one-cond}
\begin{eqnarray}
(A_L)_{nn}(m_n^2) + m_n^2 \left. \frac{\dd}{\dd p^2}
\left( (A_L)_{nn}(p^2) 
+ (A_R)_{nn}(p^2) \right) \right|_{p^2=m_n^2}
&& \nonumber \\ 
- m_n \left. \frac{\dd}{\dd p^2}
\left( (B_L)_{nn}(p^2) +(B_R)_{nn}(p^2) \right) \right|_{p^2=m_n^2} 
&=& 1,
\\ 
(A_R)_{nn}(m_n^2) + m_n^2 \left. \frac{\dd}{\dd p^2}
\left( (A_L)_{nn}(p^2) 
+ (A_R)_{nn}(p^2) \right) \right|_{p^2=m_n^2}
&& \nonumber \\ 
- m_n \left. \frac{\dd}{\dd p^2}
\left( (B_L)_{nn}(p^2) +(B_R)_{nn}(p^2) \right) \right|_{p^2=m_n^2} 
&=& 1.
\end{eqnarray}
\end{subequations}
Note that in these two equations the parts with the derivatives are identical,
therefore, we are lead to the consistency condition
\begin{equation}\label{Lnn=Rnn}
(A_L)_{nn}(m_n^2) = (A_R)_{nn}(m_n^2).
\end{equation}
Plugging this into equation~(\ref{final-c1}), the further consistency condition
\begin{equation}
(B_L)_{nn}(m_n^2) = (B_R)_{nn}(m_n^2)
\end{equation}
follows.

\subsection{Renormalization and parameter counting}
\label{rapc}
Turning to the counting of the number of renormalization conditions, 
we note that 
the relations analogous to the ones of equation~(\ref{AB}) are now
\begin{equation}\label{ABLR}
A_L^\dagger = A_L, \quad A_R^\dagger = A_R, \quad B_L^\dagger = B_R.
\end{equation}
These relations cut the number of independent relations in
equation~(\ref{final-c1}) in half and it suffices to consider only 
equation~(\ref{final-c1-a}). Written in indices $i,j$, this equation reads
\begin{equation}\label{ij-without}
(B_L)_{ij}(m_j^2) =  m_j (A_R)_{ij}(m_j^2), \quad 
(B_R)_{ij}(m_j^2) =  m_j (A_L)_{ij}(m_j^2). 
\end{equation}
In counting the number of independent on-shell conditions
we proceed as in section~\ref{rp-parityconservation}. 
We first consider $i \neq j$. The relations in equation~(\ref{ij-without}) 
are not symmetric under the exchange $i \leftrightarrow j$. 
Furthermore, each of the two independent relations in 
equation~(\ref{ij-without}) delivers one complex condition. Therefore, 
each pair of indices gives eight real conditions. 
For $i = j$, the problem is more subtle. We know from equation~(\ref{ABLR}) 
that 
$\mathrm{Im}\, (A_L)_{ii} = \mathrm{Im}\, (A_R)_{ii} = 0$ and 
$(B_R)_{ii}(m_i^2) = \left((B_L)_{ii}(m_i^2)\right)^*$.  
Clearly, these are no renormalization conditions. 
Discounting them, there remain the three real
renormalization conditions 
\begin{equation}\label{real}
(A_L)_{ii}(m_i^2) = (A_R)_{ii}(m_i^2), \quad
\mathrm{Im}\,(B_L)_{ii}(m_i^2) = 0, \quad
\mathrm{Re}\,(B_L)_{ii}(m_i^2) =  m_i (A_R)_{ii}(m_i^2)
\end{equation}
plus one condition containing the derivatives.
Actually, only the second and third relation in equation~(\ref{real})
follow from equation~(\ref{ij-without}) with $i=j$, whereas the first one
ensues from that equation only for $m_i \neq 0$. However, the first one also
derives from equation~(\ref{eq:residue-one-cond}) without the necessity of
assuming $m_i \neq 0$ and, because of this, the same equation  
gives only one real condition containing the derivatives. 
Thus we end up with a total of 
\begin{equation}\label{number-pv}
8\, \binom{N}{2} + 4N = 4N^2
\end{equation}
real on-shell conditions.

Next we switch to renormalization.
Equation~(\ref{invSF}) holds again, but now we have to introduce a field
strength renormalization matrix for each set of chiral fields, \textit{i.e.}
\begin{equation}
\psi^{(b)}_{iL} = \left( \zz_L \right)_{ik} \psi_{Lk}, \quad
\psi^{(b)}_{iR} = \left( \zz_R \right)_{ik} \psi_{Rk}.
\end{equation}
Then, the renormalized self-energy $\Sigma^{(r)}(p)$
is written as 
\begin{eqnarray}
\Sigma^{(r)}(p) &=& \Sigma(p) +
\left( \bone - \left( \zz_L \right)^\dagger \zz_L \right) \slashed{p}
\gamma_L + 
\left( \bone - \left( \zz_R \right)^\dagger \zz_R \right) \slashed{p}
\gamma_R 
\nonumber \\ && + 
\left( \zz_R \right)^\dagger \left( \hat m + \delta \hat m \right) \zz_L
\gamma_L +
\left( \zz_L \right)^\dagger \left( \hat m + \delta \hat m \right) \zz_R
\gamma_R- \hat m
\end{eqnarray}
with the unrenormalized self-energy $\Sigma(p)$.
In order to satisfy the $4N^2$ renormalization conditions, we have 
$\zz_L$, $\zz_R$ and $\delta \hat m$ at our disposal.
Just as in the case of parity conservation, we have to take the phase freedom 
\begin{equation}\label{phasefactors-LR}
\zz_L \to e^{i\hat \alpha} \zz_L, \quad\zz_R \to e^{i\hat \alpha} \zz_R
\end{equation}
into account, because $\Sigma^{(r)}(p)$ is invariant under this
rephasing. Note that both field strength renormalization matrices are
multiplied with the \emph{same} diagonal matrix of phase factors. 
Each field strength renormalization matrix is a general complex matrix and 
thus has $2N^2$ real parameters. Taking into account the $N$ parameters in 
$\delta \hat m$, we end up with total number of 
$(4N^2 - N) + N = 4N^2$ real parameters, which we have at our disposal for
renormalization, which exactly matches the number of independent real 
renormalization conditions in equation~(\ref{number-pv}). 

For renormalization, we decompose the bare self-energy into
\begin{equation}
\Sigma(p) = \slashed{p} \left( \Sigma^{(A)}_L(p^2) \gamma_L + 
 \Sigma^{(A)}_R(p^2) \gamma_R \right) + 
\Sigma^{(B)}(p^2) \gamma_L + \Sigma^{(B)}_R(p^2) \gamma_R,
\end{equation}
\begin{subequations}
which in turn gives
\begin{alignat}{2}
A_L &= \left( \zz_L \right)^\dagger \zz_L - \Sigma^{(A)}_L, & \quad
B_L &= \Sigma^{(B)}_L +
\left( \zz_R \right)^\dagger \left( \hat m + \delta \hat m \right) \zz_L,
\\
A_R &= \left( \zz_R \right)^\dagger \zz_R - \Sigma^{(A)}_R, & \quad
B_R &= \Sigma^{(B)}_R +
\left( \zz_L \right)^\dagger \left( \hat m + \delta \hat m \right) \zz_R.
\end{alignat}
\end{subequations}
Note that
\begin{equation}\label{sigma-h}
\left( \Sigma^{(A)}_L \right)^\dagger = \Sigma^{(A)}_L, \quad
\left( \Sigma^{(A)}_R \right)^\dagger = \Sigma^{(A)}_R, \quad
\left( \Sigma^{(B)}_L \right)^\dagger = \Sigma^{(B)}_R 
\end{equation}
due to the discussion in appendix~\ref{dispersive}.

Let us now perform explicit renormalization at the lowest 
non-trivial order. To this end, we assume that $\hat m$ is diagonal, with the
pole masses on its diagonal. Then $\delta \hat m$ is diagonal too. 
We write 
$\zz_L = \bone + \frac{1}{2} z_L$ and
$\zz_R = \bone + \frac{1}{2} z_R$
and obtain 
\begin{subequations}
\begin{align}
A_{L} &= \bone - \Sigma^{(A)}_{L} + \frac{1}{2}z_{L} + \frac{1}{2}z_{L}^\dagger,\\
A_{R} &= \bone - \Sigma^{(A)}_{R} + \frac{1}{2}z_{R} + \frac{1}{2}z_{R}^\dagger,
\end{align}
\end{subequations}
and
\begin{subequations}
\begin{align}
B_{L} &= \hat{m} + \delta \hat{m} + \Sigma^{(B)}_{L} + 
\frac{1}{2} \hat{m} z_{L} + \frac{1}{2}  z_{R}^\dagger \hat{m}, \\
B_{R} &= \hat{m} + \delta \hat{m} + \Sigma^{(B)}_{R} + 
\frac{1}{2} \hat{m} z_{R} + \frac{1}{2} z_{L}^\dagger\hat{m}.
\end{align}
\end{subequations}
With this identification, it is straightforward to compute $(z_L)_{ij}$ and
$(z_R)_{ij}$ for $i \neq j$ from equation~(\ref{final-c1-a}).
The result is (see, for instance,~\cite{denner,grimus})
\begin{subequations}\label{eq:FSRC-offdiag-ver1}
\begin{eqnarray}\label{zL}
\lefteqn{\frac{1}{2}(z_L)_{ij} =} \\
&& -\frac{1}{m_i^2-m_j^2} 
\left[ m_j^2 \left( \Sigma^{(A)}_L \right)_{ij} + 
m_i m_j \left( \Sigma^{(A)}_R \right)_{ij} + 
m_j \left( \Sigma^{(B)}_R \right)_{ij} +
m_i \left( \Sigma^{(B)}_L \right)_{ij} \right]_{p^2 = m_j^2}, 
\nonumber
\\ \label{zR}
\lefteqn{\frac{1}{2}(z_R)_{ij} =} \\ 
&& -\frac{1}{m_i^2-m_j^2}
\left[ m_i m_j \left( \Sigma^{(A)}_L \right)_{ij} + 
m_j^2 \left( \Sigma^{(A)}_R \right)_{ij} + 
m_i \left( \Sigma^{(B)}_R \right)_{ij} +
m_j \left( \Sigma^{(B)}_L \right)_{ij} \right]_{p^2 = m_j^2}.
\nonumber
\end{eqnarray}
\end{subequations}

For $i=j$, we have the three conditions in equation~(\ref{real}) plus one of
the conditions of equation~(\ref{eq:residue-one-cond}), \textit{i.e.}\ there
are four equations for five unknowns, which are
$\mathrm{Re}\,(z_L)_{ii}$, $\mathrm{Re}\,(z_R)_{ii}$,  
$\mathrm{Im}\,(z_L)_{ii}$, $\mathrm{Im}\,(z_R)_{ii}$ and $\delta m_i$. This
reflects the rephasing invariance of equation~(\ref{phasefactors-LR}).
To solve for $\delta m_i$, the conditions in
equation~(\ref{real}) are sufficient, leading to the result
\begin{equation}\label{eq:mass-counter}
2\,\delta \hat{m}_{ii} = 
-m_i \left( (\Sigma^{(A)}_{L})_{ii}(m_i^2) + (\Sigma^{(A)}_{R})_{ii}(m_i^2)
\right)  
-\left( (\Sigma^{(B)}_{L})_{ii}(m_i^2) + (\Sigma^{(B)}_{R})_{ii}(m_i^2) \right). 
\end{equation}
Note that, due to equation~(\ref{sigma-h}), the second expression on the
right-hand side is real.
The remaining task is the determination of the diagonal entries of the field
strength renormalization constants. 
The real parts can be completely fixed:
\begin{subequations}
\begin{align}
\mathrm{Re}(z_L)_{ii} = (\Sigma^{(A)}_L)_{ii}(m_i^2) &+ m_i^2
\left. \frac{\dd}{\dd p^2}
\left( (\Sigma^{(A)}_L)_{ii}(p^2) + (\Sigma^{(A)}_R)_{ii}(p^2)
  \right)\right|_{p^2=m_i^2} \nonumber \\ 
   &+ m_i \left. \frac{\dd}{\dd p^2}\left( (\Sigma^{(B)}_L)_{ii}(p^2) +
  (\Sigma^{(B)}_R)_{ii}(p^2) \right)\right|_{p^2=m_i^2},
\label{ReziiL} \\ 
  \mathrm{Re}(z_R)_{ii}= (\Sigma^{(A)}_R)_{ii}(m_i^2) &+ m_i^2 \left. \frac{\dd}{\dd
    p^2}\left( (\Sigma^{(A)}_L)_{ii}(p^2) + (\Sigma^{(A)}_R)_{ii}(p^2)
  \right)\right|_{p^2=m_n^2} \nonumber \\ 
   &+ m_i \left. \frac{\dd}{\dd p^2}\left( (\Sigma^{(B)}_L)_{ii}(p^2) +
  (\Sigma^{(B)}_R)_{ii}(p^2) \right)\right|_{p^2=m_i^2}. 
\label{ReziiR}
\end{align}
\end{subequations}
Only $\mathrm{Im}\,(B_L)_{ii}(m_i^2) = 0$ in equation~(\ref{real}) involves a
non-trivial imaginary part, from which we deduce~\cite{grimus} 
\begin{equation}\label{Imzii}
m_i\, \mathrm{Im}\, (z_L - z_R)_{ii} = 
-2\,\mathrm{Im}\,(\Sigma^{(B)}_L)_{ii}(m_i^2) =
2\,\mathrm{Im}\,(\Sigma^{(B)}_R)_{ii}(m_i^2). 
\end{equation}
The latter equality results from equation~(\ref{sigma-h}).

\subsection{Comparison with Aoki \textit{et al.}}
In~\cite{aoki}, the inverse
propagator is parameterized as 
\begin{equation}
S^{-1} = K_1 \bone + K_5 \gamma_5 + K_\gamma \slashed{p} + K_{5\gamma}
\slashed{p} \gamma_5.
\end{equation}
Comparison with equation~(\ref{Sinv}) leads to the translation table
\begin{equation}
\renewcommand{\arraystretch}{1.2}
\begin{array}{lcr}
K_1 &=& -\frac{1}{2} \left( B_L + B_R \right), \\
K_5 &=&  \frac{1}{2} \left( B_L - B_R \right), \\
K_\gamma &=& \frac{1}{2} \left( A_L + A_R \right), \\
K_{5\gamma} &=& -\frac{1}{2} \left( A_L - A_R \right). 
\end{array}
\end{equation}
Inverting it, we have
\begin{equation}
\renewcommand{\arraystretch}{1.2}
\begin{array}{lcr}
A_L &=& K_\gamma - K_{5\gamma}, \\
A_R &=& K_\gamma + K_{5\gamma}, \\
B_L &=& -K_1 + K_5, \\
B_R &=& -K_1 - K_5.
\end{array}
\end{equation}
With this information, equation~(\ref{final-c1}) is translated into
\begin{equation}
\begin{array}{ccc}
(K_1)_{in}(m_n^2) + m_n (K_\gamma)_{in}(m_n^2) = 0, &
(K_5)_{in}(m_n^2) - m_n (K_{5\gamma})_{in}(m_n^2) = 0  & \forall i, 
\\
(K_1)_{nj}(m_n^2) + m_n (K_\gamma)_{nj}(m_n^2) = 0, &
(K_5)_{nj}(m_n^2) + m_n (K_{5\gamma})_{nj}(m_n^2) = 0 & \forall j.
\end{array}
\end{equation}
For $i=j=n$, we find 
\begin{subequations}
\begin{eqnarray}
&&
(K_1)_{nn}(m_n^2) + m_n (K_\gamma)_{nn}(m_n^2) = 0, \quad 
(K_5)_{nn}(m_n^2) = 0, \quad 
(K_{5\gamma})_{nn}(m_n^2) = 0, \\
&&
(K_\gamma)_{nn}(m_n^2) + 
2 m_n^2 \left. \frac{\dd K_\gamma(p^2)}{\dd p^2} \right|_{p^2 = m_n^2} +
2 m_n \left. \frac{\dd K_1(p^2)}{\dd p^2} \right|_{p^2 = m_n^2} = 1.
\end{eqnarray}
\end{subequations}
Therefore, we have full agreement with the on-shell conditions in~\cite{aoki}.

\subsection{On-shell renormalization of Majorana fermions}
In the case of Majorana fermions, the propagator matrix is severely restricted
because each fermion field is its own charge-conjugate field---see
appendix~\ref{Majorana condition}. For instance, 
the bare fields can represented as 
\begin{equation}
\psi^{(b)}_i = \psi^{(b)}_{Li} + \left( \psi^{(b)}_{Li} \right)^c,
\end{equation}
where the superscript $c$ denotes charge conjugation.
This formula expresses the fact that for each field only one chiral component is
independent. Therefore, 
\begin{equation}
\zz_R = \left( \zz_L \right)^*
\end{equation}
holds~\cite{kniehl1996}. The freedom of rephasing expressed by 
equation~(\ref{phasefactors-LR}) is lost, except for those indices $i$ for
which $m_i + \delta m_i = 0$. 
However, in the following discussion, we will exclude such cases. Therefore, 
real parameters we have at disposal for renormalization are those in
$\zz_L$ and $\delta \hat m$. This amounts to $2N^2 + N$ parameters. 

The Majorana condition 
\begin{equation}
S^{-1}(p) = C \left( S^{-1}(-p) \right)^T C^{-1}
\end{equation}
on the propagator matrix is derived
in appendix~\ref{Majorana condition}. Applying
this condition to the inverse propagator as parameterized in
equation~(\ref{Sinv}), one readily finds 
\begin{equation}
A_L^T = A_R, \quad B_L^T = B_L, \quad B_R^T = B_R.
\end{equation}
These conditions hold, in the Majorana case, in addition to those of
equation~(\ref{ABLR}). Therefore, in summary we have here
\begin{equation}\label{AB-M}
A_R = A_L^* \;\; \mbox{with} \;\; A_L^\dagger = A_L
\quad \mbox{and} \quad 
B_R = B_L^* \;\; \mbox{with} \;\; B_L^T = B_L.
\end{equation}
It is important not to confuse these relations with renormalization
conditions. 

In order to count the number of the latter ones in the case of
Majorana fermions, we first consider $i \neq j$. As discussed in
section~\ref{rapc}, it suffices to consider equation~(\ref{ij-without}).
However, using equation~(\ref{AB-M}) to eliminate $B_R$ and $A_L$ in the
second relation of this equation, we find that in the Majorana case 
the two relations are equivalent. Thus, for $i \neq j$ there are 
$4{N \choose 2}$ independent real renormalization conditions.
For $i = j$, we reconsider equation~(\ref{real}). The first relation in this
equation now follows from equation~(\ref{AB-M}) and must not be considered as a
renormalization condition. Therefore, in summary, together with the residue
condition of 
equation~(\ref{eq:residue-one-cond}), we have 
\begin{equation}
4 {N \choose 2} + 3N = 2N^2 + N
\end{equation}
renormalization conditions, which matches the number of 
independent renormalization constants determined above. 

Equation~(\ref{AB-M}) holds for the unrenormalized self-energies as well, in
particular, the relations
\begin{equation}
\Sigma_R^{(A)} = \left( \Sigma_L^{(A)} \right)^*
\quad \mbox{and} \quad
\Sigma_R^{(B)} = \left( \Sigma_L^{(B)} \right)^*
\end{equation}
are valid. Using these to eliminate the self-energy parts 
$\Sigma_R^{(A)}$ and $\Sigma_R^{(B)}$ in the one-loop renormalization
conditions of section~\ref{rapc}, it is easy to see their 
compatibility with the Majorana condition. Indeed, one finds that
equation~(\ref{zR}) is the complex conjugate of equation~(\ref{zL}) and the
same is true for equations~(\ref{ReziiL}) and~(\ref{ReziiR}).
Equation~(\ref{Imzii}) now reads 
\begin{equation}
m_i\, \mathrm{Im}\, (z_L)_{ii} = 
-\mathrm{Im}\,(\Sigma^{(B)}_L)_{ii}(m_i^2)
\end{equation}
because of $\mathrm{Im}\, (z_R)_{ii} = -\mathrm{Im}\, (z_L)_{ii}$.
As discussed above, there is no phase freedom in the Majorana case.

\section{Summary}
\label{summary}
In this review we have given a very explicit presentation of on-shell
renormalization. In particular, we have taken pains to dispel any unclear
point in the derivation of the on-shell renormalization conditions imposed on
the inverse propagator matrix. We have extensively discussed mixing of $N$
fields in the case of real scalar fields and fermion fields in 
parity-conserving and parity-violating theories. We have also distinguished
between Dirac and Majorana fermions and described how the
renormalization scheme gets modified in the latter case as compared to the
more familiar Dirac case. We have not treated here mixing of
complex scalar fields, but with the methods explained in this review this 
should pose no problem. Moreover, we have omitted mixing of vector fields, in
order to avoid the complications of gauge theories---for photon--$Z$ boson
mixing we refer the reader to~\cite{sirlin,hollik}, for instance. 
The main motivation for this review originates from extensions of the Standard
Model in the fermion and scalar sector, with special focus on Majorana
neutrinos. 
For self-consistency and because of the important role in our discussion, 
we have also supplied a derivation of the emergence of dispersive and absorptive
parts in the propagator matrix and a derivation of the restrictions on the
propagator matrix in the case of Majorana fermions.

\section*{Acknowledgments}
M.L. is supported by the Austrian Science Fund (FWF), Project
No.\ P28085-N27 and in part by the FWF Doctoral Program
No.\ W1252-N27 Particles and Interactions. 
Both authors thank H.~Neufeld for useful discussions.

\appendix

\setcounter{equation}{0}
\renewcommand{\theequation}{A\arabic{equation}}
\section{Dispersive and absorptive parts in the propagator}
\label{dispersive}
In this review we stick to hermitian counterterms in the Lagrangian.
In the present appendix, for the sake of completeness, we want to demonstrate
that the dispersive parts of the 
propagator matrix fulfill hermiticity conditions matching those of the
counterterms. This is a necessary prerequisite for imposing on-shell
conditions. Dispersive and absorptive parts arise
through the well-known relation in distribution theory 
\begin{equation}\label{sp}
\frac{1}{p^2 - \mu^2 + i\epsilon}  = \mbox{P} \frac{1}{p^2 - \mu^2} - 
i\pi \delta(p^2 - \mu^2),
\end{equation}
from the principle value and the delta function, respectively. This can be
verified quite easily in the K\"all\'en--Lehmann representation of the
propagator matrix. By and large we follow the derivation in~\cite{zuber}.

\subsection{Real scalar fields}
\label{sec:scalar-prop-symmetry}
The total propagator matrix is defined by
\begin{equation}\label{d'f}
i \left( \Delta'(x-y) \right)_{ij} = 
\langle 0 | \mathrm{T} \varphi_i(x) \varphi_j(y) | 0 \rangle,
\end{equation}
where T denotes time ordering. The propagator $\Delta'$ 
could be renormalized or unrenormalized. 
Inserting a complete system of four-momentum eigenstates and exploiting energy
momentum conservation leads to
\begin{eqnarray}
i \left( \Delta'(x-y) \right)_{ij} &=& 
\sum_n \left\{ 
\Theta(x^0-y^0) \,
\langle 0 | \varphi_i(0) | n \rangle 
\langle n | \varphi_j(0) | 0 \rangle e^{-ip_n \cdot (x-y)} \right.
\nonumber \\ &&
\left. + 
\Theta(y^0-x^0) \,
\langle 0 | \varphi_j(0) | n \rangle 
\langle n | \varphi_i(0) | 0 \rangle e^{ip_n \cdot (x-y)} \right\}.
\end{eqnarray}
It is then useful to define the density 
\begin{equation}
(2\pi)^3 \sum_n \delta^{(4)}(q-p_n) 
\langle 0 | \varphi_i(0) | n \rangle 
\langle n | \varphi_j(0) | 0 \rangle 
\equiv \rho_{ij}(q^2) \Theta(q^0),
\end{equation}
where the Heaviside function $\Theta$ indicates that only $q^0 \geq 0$ gives a
contribution.

Next we invoke CPT invariance, which holds in any local, Lorentz-invariant
field theory. The transformation of the real scalar fields is given by 
\begin{equation}\label{cpt-s}
\cpt \varphi_i(x) \cpt^{-1} = \varphi_i(-x).
\end{equation}
Since $\cpt$ is an antiunitary operator and $\cpt | 0 \rangle = | 0 \rangle$,
we have the relation
\begin{equation}\label{cpt0}
\langle 0 | \varphi_i(0) | n \rangle = 
\big( \langle 0 | \varphi_i(0) \cpt | n \rangle \big)^*.
\end{equation}
Clearly, if $| n \rangle$ is a complete system, then 
$\cpt | n \rangle$ is a complete system as well. Application 
of equation~(\ref{cpt0}) to $\rho_{ij}(q^2)$ gives the two relations 
\begin{equation}
\rho_{ij}(q^2) = \left( \rho_{ij}(q^2) \right)^* = \rho_{ji}(q^2).
\end{equation}
The second one allows us to write the propagator matrix as
\begin{equation}
i\Delta'(x-y) = \frac{1}{(2\pi)^3} \int \dd^4 q\, \Theta(q^0) \rho(q^2) 
\left( \Theta(x^0 - y^0) e^{-iq \cdot (x-y)} + 
\Theta(y^0 - x^0) e^{iq \cdot (x-y)} \right). 
\end{equation}
By insertion of $1 = \int_0^\infty \dd \mu^2\, \delta(q^2 - \mu^2)$
one ends up with the K\"all\'en--Lehmann representation
\begin{equation}
i\Delta'(x-y) = 
i \int_0^\infty \dd \mu^2 \rho(\mu^2) \Delta_F(x-y;\mu),
\end{equation}
where 
\begin{equation}
\Delta_F(x-y;\mu) = -i \int \frac{\dd^3 q}%
{(2\pi)^3 2 \sqrt{\mu^2 + {\vec q}^{\,2}}} \left(
\Theta(x^0-y^0) e^{-i q \cdot (x-y)} + 
\Theta(y^0-x^0) e^{i q \cdot (x-y)} \right)
\end{equation}
is the free Feynman propagator of a single real scalar
field with mass $\mu$ and $q^2 = \mu^2$. 
In momentum space, this formula reads
\begin{equation}
\Delta'(p^2) = 
\int_0^\infty \dd \mu^2 \rho(\mu^2) \frac{1}{p^2 - \mu^2 + i\epsilon}.
\end{equation}

If we identify $\Delta'(p^2)$ with the renormalized Feynman propagator, 
we see that $\Delta_{ij}(p^2)$ is real and symmetric in the region of
$p^2$ where the absorptive part vanishes. This justifies the assumption made
in section~\ref{scalar-parameters}.

\subsection{Fermion fields}\label{sec:fermion-prop-symmetry}
In the fermionic case we have
\begin{eqnarray}\label{SF'}
i \left( S'(x-y) \right)_{ia,jb} &=& 
\langle 0 | \mathrm{T} \psi_{ia}(x) \bar\psi_{jb}(y) | 0 \rangle 
\\ \nonumber & \equiv &
\Theta(x^0-y^0) \langle 0 | \psi_{ia}(x) \bar\psi_{jb}(y) | 0 \rangle -
\Theta(y^0-x^0) \langle 0 |  \bar\psi_{jb}(y)\psi_{ia}(x) | 0 \rangle,
\end{eqnarray}
where for the sake of clarity we have also introduced Dirac indices
$a,b$ for each field. With the first term on the right-hand side of
equation~(\ref{SF'}) one can, just as in the scalar case, define a density:
\begin{equation}
(2\pi)^3 \sum_n \delta^{(4)}(q - p_n)
\langle 0 | \psi_{ia}(0) | n \rangle 
\langle n | \bar\psi_{jb}(0) | 0 \rangle \equiv \rho_{ia,jb}(q) \Theta(q^0).
\end{equation}
Lorentz-invariance permits the decomposition 
\begin{equation}\label{rho-f}
\rho(q) = \slashed{q} \left( c_L(q^2) \gamma_L + c_R(q^2) \gamma_R \right) +
d_L(q^2) \gamma_L + d_R(q^2) \gamma_R.
\end{equation}
In this formula, $c_{L,R}$, $d_{L,R}$ are $N \times N$ matrices in
family space, while $\slashed{q} \gamma_{L,R}$, $\gamma_{L,R}$ carry the Dirac
indices.  

In order to relate the second term on the right-hand side of
equation~(\ref{SF'}) to the density $\rho$, CPT-invariance has to be
invoked~\cite{zuber}. To be as explicit as possible, we present all our
conventions. We use the defining relation
\begin{equation}
C^{-1} \gamma_\mu C = -\gamma_\mu^T
\end{equation}
for the charge conjugation matrix, whence 
$C^T = -C$ ensues. As for the Dirac gamma matrices, we assume the hermiticity
conditions $\gamma_0^\dagger = \gamma_0$ and 
$\gamma_i^\dagger = -\gamma_i$ for $i=1,2,3$. In this basis, 
$C^\dagger = C^{-1}$ without loss of generality. 
Charge conjugation, parity and time reversal transformation for fermion fields
are formulated as
\begin{equation}
\mathcal{C} \psi_i(x) \mathcal{C}^{-1} = C \gamma_0^T \psi_i^*(x),
\;\;
\mathcal{P} \psi_i(x) \mathcal{P}^{-1} = \gamma_0 \psi_i(x^0,-\vec x\,),
\;\; 
\mathcal{T} \psi_i(x) \mathcal{T}^{-1} = 
i C^{-1} \gamma_5 \psi_i(-x^0,\vec x\,),
\end{equation}
respectively. Combination of the three discrete transformations gives the CPT
transformation property~\cite{zuber,rebelo} 
\begin{equation}\label{cpt-f}
\cpt \psi_i(x) \cpt^{-1} = -i \gamma_5^T \psi_i^*(-x)
\quad \mbox{and} \quad
\cpt \bar\psi_i(x) \cpt^{-1} = i \psi_i^T(-x) (\gamma_5 \gamma_0)^*,
\end{equation}
where the overall sign is convention, but the $i$ is necessary in Yukawa
couplings.\footnote{
The reason is that the transformation~(\ref{cpt-s}) together
with~(\ref{cpt-f}) leaves the most general Yukawa coupling 
\begin{equation}\nonumber
\sum_{r,s,k} \Gamma_{rs}^k \psi_r^T C^{-1} \psi_s\, \varphi_k + \mbox{H.c.}
\quad \mbox{with} \quad \Gamma_{rs}^k = \Gamma_{sr}^k \;\forall \; r,s
\end{equation}
invariant; in this Yukawa coupling, all fermion fields are, for instance,
left-chiral. Therefore, equations~(\ref{cpt-s}) and~(\ref{cpt-f}) together 
form a true CPT transformation.}

Now we apply equation~(\ref{cpt-f}) to the second term on the right-hand side
of the propagator~(\ref{SF'}). A straightforward computation, taking into
account antiunitary of $\cpt$ and the above conventions for $C$ and the gamma
matrices, leads to the simple result~\cite{zuber}
\begin{equation}\label{term2}
\langle 0 | \bar \psi_{jb}(y) \psi_{ia}(x) | 0 \rangle =
-\left( \gamma_5 \right)_{ad} 
\langle 0 | \psi_{id}(-x) \bar\psi_{jc}(-y) | 0 \rangle
\left( \gamma_5 \right)_{cb}.
\end{equation}
The minus sign is used to cancel the minus from the time ordering when 
$y^0 > x^0$. Thus, for the contribution to the propagator of
equation~(\ref{term2}), we obtain the density
\begin{equation}
\gamma_5 \rho(q) \gamma_5 = 
-\slashed{q} \left( c_L(q^2) \gamma_L + c_R(q^2) \gamma_R \right) +
d_L(q^2) \gamma_L + d_R(q^2) \gamma_R,
\end{equation}
with $\rho(q)$ being identical to that of equation~(\ref{rho-f}).

The remaining steps are the same as in the scalar case. We finally arrive at
\begin{equation}
iS'(x-y) = i\int_0^\infty \dd \mu^2 \left[ 
i \slashed{\partial}_x \left( c_L(\mu^2) \gamma_L + c_R(\mu^2) \gamma_R \right)
+ d_L(\mu^2) \gamma_L + d_R(\mu^2) \gamma_R \right] \Delta_F(x-y;\mu),
\end{equation}
where the subscript $x$ indicates derivative with respect to $x$.
In momentum space the result is
\begin{equation}
S'(p) = \int_0^\infty \dd \mu^2 \left[ 
\slashed{p} \left( c_L(\mu^2) \gamma_L + c_R(\mu^2) \gamma_R \right)
+ d_L(\mu^2) + d_R(\mu^2) \right] \frac{1}{p^2 - \mu^2 + i\epsilon}.
\end{equation}

From the definition of the density $\rho(q)$, equation~(\ref{rho-f}), the
following property is easy to prove:
\begin{equation}
\gamma_0 \rho^\dagger(q) \gamma_0 = \rho(q) 
\quad \Rightarrow \quad
c_L^\dagger (q^2) = c_L(q^2), \;
c_R^\dagger (q^2) = c_R(q^2), \;
d_L^\dagger (q^2) = d_R(q^2).
\end{equation}
Transferring the latter relations to the dispersive part of $S'$, we obtain
\begin{equation}
\gamma_0 (S'(p))_\mathrm{disp}^\dagger \gamma_0 = 
(S'(p))_\mathrm{disp}.
\end{equation}
This is the justification of equation~(\ref{Scond}).

\setcounter{equation}{0}
\renewcommand{\theequation}{B\arabic{equation}}

\section{Computational details}
\label{computational}
\subsection{Theories with parity conservation}
\label{app-pc}
We first consider the second relation in each line of equation~(\ref{ss}). The
expansion in $\ve$ furnishes
\begin{subequations}\label{abcd0}
\begin{eqnarray}
\frac{\delta_{in}}{\ve} \left( B^{(0)}_{nj} -  m_n A^{(0)}_{nj} \right) +
\delta_{in} \left( B^{(1)}_{nj} - m_n A^{(1)}_{nj} \right) + 
C^{(0)}_{ik} B^{(0)}_{kj}\, + D^{(0)}_{ik} A^{(0)}_{kj} + \coe = 0,
\\
\left( B^{(0)}_{in} -  m_n A^{(0)}_{in} \right) \frac{\delta_{nj}}{\ve} +
\left( B^{(1)}_{in} - m_n A^{(1)}_{in} \right) \delta_{nj} + 
B^{(0)}_{ik} C^{(0)}_{kj}\, + A^{(0)}_{ik} D^{(0)}_{kj} + \coe = 0.
\end{eqnarray}
\end{subequations}
Removing the singularity, we obtain the conditions
\begin{equation}\label{fc1}
B^{(0)}_{in} =  m_n A^{(0)}_{in} \;\; \forall\; i = 1,\ldots,N
\quad \mbox{and} \quad
B^{(0)}_{nj} =  m_n A^{(0)}_{nj} \;\; \forall\; j = 1,\ldots,N.
\end{equation}
Next we consider the first relation in each line of equation~(\ref{ss}).
Taking into account $p^2 = \ve + m_n^2$, we find
\begin{subequations}
\begin{eqnarray}\label{abcd}
-\frac{\delta_{in} m_n}{\ve} \left( B^{(0)}_{nj} -  m_n A^{(0)}_{nj} \right) 
&& \nonumber \\
+\,
\delta_{in} \left( A^{(0)}_{nj} + m_n^2  A^{(1)}_{nj} - m_n B^{(1)}_{nj}
\right) + 
m_n^2 C^{(0)}_{ik} A^{(0)}_{kj}\, + D^{(0)}_{ik} B^{(0)}_{kj} + \coe 
&=& \delta_{ij},
\\
-\left( B^{(0)}_{in} -  m_n A^{(0)}_{in} \right) \frac{m_n \delta_{nj}}{\ve} 
&& \nonumber \\ 
+\,
\left( A^{(0)}_{in} + m_n^2 A^{(1)}_{in} - m_n B^{(1)}_{in} \right) \delta_{nj} 
+
m_n^2 A^{(0)}_{ik} C^{(0)}_{kj} + B^{(0)}_{ik} D^{(0)}_{kj}  + \coe &=& \delta_{ij}.
\end{eqnarray}
\end{subequations}
The singularity in these relations leads again to equation~(\ref{fc1}). 

Up to now we have only considered the singular terms. In order to obtain a
renormalization condition from the terms of order $\ve^0$, we choose the
indices $i=j=n$. In this case, equation~(\ref{abcd}) leads to
\begin{subequations}
\begin{eqnarray}
A^{(0)}_{nn} + m_n^2  A^{(1)}_{nn} - m_n B^{(1)}_{nn} + 
m_n^2 C^{(0)}_{nk} A^{(0)}_{kn} + D^{(0)}_{nk} B^{(0)}_{kn} &=& 1,
\\
A^{(0)}_{nn} + m_n^2 A^{(1)}_{nn} - m_n B^{(1)}_{nn} +
m_n^2 A^{(0)}_{nk} C^{(0)}_{kn} + B^{(0)}_{nk} D^{(0)}_{kn} &=& 1.
\end{eqnarray}
\end{subequations}
Invoking equation~(\ref{fc1}), we rewrite this equation as
\begin{subequations}\label{abcd1}
\begin{eqnarray}
A^{(0)}_{nn} + m_n^2  A^{(1)}_{nn} - m_n B^{(1)}_{nn} + 
\left( m_n C^{(0)}_{nk} + D^{(0)}_{nk} \right) A^{(0)}_{kn} m_n &=& 1,
\\
A^{(0)}_{nn} + m_n^2 A^{(1)}_{nn} - m_n B^{(1)}_{nn} +
m_n A^{(0)}_{nk} \left( m_n C^{(0)}_{kn} + D^{(0)}_{kn} \right) &=& 1.
\end{eqnarray}
\end{subequations}
In the same way equation~(\ref{abcd0}) gives
\begin{subequations}
\begin{eqnarray}
B^{(1)}_{nn} - m_n A^{(1)}_{nn} + 
\left( m_n C^{(0)}_{nk} + D^{(0)}_{nk} \right) A^{(0)}_{kn} &=& 0,
\\
B^{(1)}_{nn} - m_n A^{(1)}_{nn} + 
A^{(0)}_{nk} \left( m_n C^{(0)}_{kn} + D^{(0)}_{kn} \right) &=& 0.
\end{eqnarray}
\end{subequations}
These relations allow us to eliminate in equation~(\ref{abcd1}) the terms in
parentheses. We obtain one further renormalization condition
\begin{equation}\label{fc2}
A^{(0)}_{nn} + 2m_n^2 A^{(1)}_{nn} - 2m_n B^{(1)}_{nn} = 1.
\end{equation}

\renewcommand{\theequation}{B\arabic{equation}}
\subsection{Theories without parity conservation}
\label{app-without}
Inserting the expansions~(\ref{prop-exp}) and~(\ref{invprop-exp}) of
$S$ and $S^{-1}$, respectively,  
into equation~(\ref{eq:prop-times-inverseprop}), right column, leads to
\begin{subequations} \label{eq:zero-conditions}
\begin{eqnarray}
\frac{\delta_{in}}{\ve} \left( (B^{(0)}_L)_{nj} -  m_n (A^{(0)}_L)_{nj}
\right) +
\delta_{in} \left( (B^{(1)}_L)_{nj} - m_n (A^{(1)}_L)_{nj} \right) 
&& \nonumber \\ + 
(C^{(0)}_L)_{ik} (B^{(0)}_L)_{kj} + (D^{(0)}_R)_{ik} (A^{(0)}_L)_{kj} + \coe 
&=& 0,
\\
\frac{\delta_{in}}{\ve} \left( (B^{(0)}_R)_{nj} -  m_n (A^{(0)}_R)_{nj}
\right) +
\delta_{in} \left( (B^{(1)}_R)_{nj} - m_n (A^{(1)}_R)_{nj} \right) 
&& \nonumber \\ + 
(C^{(0)}_R)_{ik} (B^{(0)}_R)_{kj} + (D^{(0)}_L)_{ik} (A^{(0)}_R)_{kj} + \coe 
&=& 0,
\\
\left( (B^{(0)}_R)_{in} -  m_n (A^{(0)}_L)_{in} \right) \frac{\delta_{nj}}{\ve} +
\left( (B^{(1)}_R)_{in} - m_n (A^{(1)}_L)_{in} \right) \delta_{nj} 
&& \nonumber \\ + 
(B^{(0)}_R)_{ik} (C^{(0)}_L)_{kj} + (A^{(0)}_L)_{ik} (D^{(0)}_L)_{kj} + \coe 
&=& 0,
\\
\left( (B^{(0)}_L)_{in} -  m_n (A^{(0)}_R)_{in} \right) \frac{\delta_{nj}}{\ve} +
\left( (B^{(1)}_L)_{in} - m_n (A^{(1)}_R)_{in} \right) \delta_{nj} 
&& \nonumber \\ + 
(B^{(0)}_L)_{ik} (C^{(0)}_R)_{kj} + (A^{(0)}_R)_{ik} (D^{(0)}_R)_{kj} + \coe 
&=& 0.
\end{eqnarray}
\end{subequations}
The poles in $\ve$ must vanish, which gives the conditions
\begin{subequations}\label{abab}
\begin{eqnarray}
(B^{(0)}_L)_{in} =  m_n (A^{(0)}_R)_{in}, \quad 
(B^{(0)}_R)_{in} =  m_n (A^{(0)}_L)_{in} 
&& \forall\; i = 1,\ldots,N,
\\
(B^{(0)}_L)_{nj} =  m_n (A^{(0)}_L)_{nj}, \quad
(B^{(0)}_R)_{nj} =  m_n (A^{(0)}_R)_{nj} 
&& \forall\; j = 1,\ldots,N.
\end{eqnarray}
\end{subequations}
Now we can insert the same expansions into
equation~(\ref{eq:prop-times-inverseprop}), left column, set $p^2= \ve +
m_n^2$ and arrive at 
\begin{subequations}
\label{LRdelta_ij}
\begin{eqnarray}
&&
-\frac{\delta_{in} m_n}{\ve} \left( (B^{(0)}_L)_{nj} -  m_n (A^{(0)}_L)_{nj}
\right) + \delta_{in} \left( (A^{(0)}_L)_{nj} + m_n^2  (A^{(1)}_L)_{nj} 
- m_n (B^{(1)}_L)_{nj} \right)
\nonumber \\
&& + \, 
m_n^2 (C^{(0)}_R)_{ik} (A^{(0)}_L)_{kj} + (D^{(0)}_L)_{ik} (B^{(0)}_L)_{kj} + \coe 
= \delta_{ij},
\\
&&
-\frac{\delta_{in} m_n}{\ve} \left( (B^{(0)}_R)_{nj} -  m_n (A^{(0)}_R)_{nj}
\right) + \delta_{in} \left( (A^{(0)}_R)_{nj} + m_n^2  (A^{(1)}_R)_{nj} 
- m_n (B^{(1)}_R)_{nj} \right)
\nonumber \\
&& + \, 
m_n^2 (C^{(0)}_L)_{ik} (A^{(0)}_R)_{kj} + (D^{(0)}_R)_{ik} (B^{(0)}_R)_{kj} + \coe 
= \delta_{ij},
\\
&& 
-\left( (B^{(0)}_L)_{in} -  m_n (A^{(0)}_R)_{in} \right) 
\frac{m_n \delta_{nj}}{\ve} 
+ \left( (A^{(0)}_R)_{in} + m_n^2 (A^{(1)}_R)_{in} - m_n (B^{(1)}_L)_{in}
\right) \delta_{nj} 
\nonumber \\ && +\,
m_n^2 (A^{(0)}_R)_{ik} (C^{(0)}_L)_{kj} + (B^{(0)}_L)_{ik} (D^{(0)}_L)_{kj}  + \coe =
\delta_{ij},
\\
&& 
-\left( (B^{(0)}_R)_{in} -  m_n (A^{(0)}_L)_{in} \right) 
\frac{m_n \delta_{nj}}{\ve} 
+ \left( (A^{(0)}_L)_{in} + m_n^2 (A^{(1)}_L)_{in} - m_n (B^{(1)}_R)_{in}
\right) \delta_{nj} 
\nonumber \\ && +\,
m_n^2 (A^{(0)}_L)_{ik} (C^{(0)}_R)_{kj} + (B^{(0)}_R)_{ik} (D^{(0)}_R)_{kj}  + \coe =
\delta_{ij}.
\end{eqnarray}
\end{subequations}
The terms with $1/\ve$ lead again to equation~(\ref{abab}).
Next, we investigate equation~(\ref{LRdelta_ij}) for $i=j=n$ at order 
$\ve^0$ and make use of equation~(\ref{abab}). 
In this way we obtain
\begin{subequations}
\label{LRnn}
\begin{eqnarray}
(A^{(0)}_L)_{nn} + m_n^2  (A^{(1)}_L)_{nn} 
- m_n (B^{(1)}_L)_{nn} +
m_n^2 (C^{(0)}_R)_{nk} (A^{(0)}_L)_{kn} + m_n (D^{(0)}_L)_{nk} (A^{(0)}_R)_{kn}
&=& 1, 
\nonumber \\ && \\
(A^{(0)}_R)_{nn} + m_n^2  (A^{(1)}_R)_{nn} 
- m_n (B^{(1)}_R)_{nn} +
m_n^2 (C^{(0)}_L)_{ik} (A^{(0)}_R)_{kj} + m_n (D^{(0)}_R)_{nk} (A^{(0)}_L)_{kn} 
&=& 1,
\nonumber \\ && \\
(A^{(0)}_R)_{nn} + m_n^2 (A^{(1)}_R)_{nn} - m_n (B^{(1)}_L)_{nn} +
m_n^2 (A^{(0)}_R)_{ik} (C^{(0)}_L)_{kj} + m_n (A^{(0)}_L)_{nk} (D^{(0)}_L)_{kn} 
&=& 1,
\nonumber \\ && \\
(A^{(0)}_L)_{nn} + m_n^2 (A^{(1)}_L)_{nn} - m_n (B^{(1)}_R)_{nn} +
m_n^2 (A^{(0)}_L)_{nk} (C^{(0)}_R)_{kn} + m_n (A^{(0)}_R)_{nk} (D^{(0)}_R)_{kn} 
&=& 1.
\nonumber \\ && 
\end{eqnarray}
\end{subequations}
Via equation~(\ref{eq:zero-conditions}) for $i=j=n$ at order $\ve^0$, 
\textit{i.e.}
\begin{subequations}
\begin{eqnarray}
(B^{(1)}_L)_{nn} - m_n (A^{(1)}_L)_{nn} + 
m_n (C^{(0)}_L)_{nk} (A^{(0)}_R)_{kn} + (D^{(0)}_R)_{nk} (A^{(0)}_L)_{kn} 
&=& 0,
\\
(B^{(1)}_R)_{nn} - m_n (A^{(1)}_R)_{nn} + 
m_n (C^{(0)}_R)_{nk} (A^{(0)}_L)_{kn} + (D^{(0)}_L)_{nk} (A^{(0)}_R)_{kn} 
&=& 0,
\\
(B^{(1)}_R)_{nn} - m_n (A^{(1)}_L)_{nn} + 
m_n (A^{(0)}_R)_{nk} (C^{(0)}_L)_{kn} + (A^{(0)}_L)_{nk} (D^{(0)}_L)_{kn} 
&=& 0,
\\
(B^{(1)}_L)_{nn} - m_n (A^{(1)}_R)_{nn} + 
m_n (A^{(0)}_L)_{nk} (C^{(0)}_R)_{kn} + (A^{(0)}_R)_{nk} (D^{(0)}_R)_{kn} 
&=& 0,
\end{eqnarray}
\end{subequations}
we can eliminate the terms with $C_{L,R}$ and $D_{L,R}$ in equation~(\ref{LRnn}).
Eventually, this leads to the final two on-shell conditions 
\begin{subequations}\label{=1}
\begin{eqnarray}
(A^{(0)}_L)_{nn} + m_n^2  \left( (A^{(1)}_L)_{nn} + (A^{(1)}_R)_{nn} \right) 
- m_n \left( (B^{(1)}_L)_{nn} +(B^{(1)}_R)_{nn} \right) &=& 1,
\\ 
(A^{(0)}_R)_{nn} + m_n^2  \left( (A^{(1)}_L)_{nn} + (A^{(1)}_R)_{nn} \right) 
- m_n \left( (B^{(1)}_L)_{nn} +(B^{(1)}_R)_{nn} \right) &=& 1.
\end{eqnarray}
\end{subequations}

\setcounter{equation}{0}
\renewcommand{\theequation}{C\arabic{equation}}
\section{Majorana condition for the propagator matrix}
\label{Majorana condition}
In the case of Majorana fields, a condition on the propagator matrix arises
from the fact that each fermion field $\psi_n$ is identical to its
charge-conjugate field $(\psi_n)^c$. Dealing with four-component spinors, this
reads
\begin{equation}\label{cc}
C \gamma_0^T \psi_n^*(x) = \psi_n(x),
\end{equation}
where $C$ is the charge-conjugation matrix.
The star means that one has to take the 
hermitian conjugate of each component
$\psi_{na}$ ($a=1,\ldots,4$) of $\psi_n$ such that $\psi_n^*$ is a column
vector. 

The starting point for the derivation of a propagator condition in the
case of Majorana fermions is the identity
\begin{equation}
\langle 0 | \mathrm{T} \psi_{ia}(x) \psi_{jb}(y) | 0 \rangle = 
-\langle 0 | \mathrm{T} \psi_{jb}(y) \psi_{ia}(x) | 0 \rangle.
\end{equation}
With $\psi_j^T = - \bar\psi_j C$ and $\psi_i = C \bar\psi_i^T$, which follow
from equation~(\ref{cc}), this identity is rewritten as
\begin{equation}
\sum_{c=1}^4\, \langle 0 | \mathrm{T} \psi_{ia}(x) \bar\psi_{jc}(y) | 0
\rangle (-C_{cb}) = 
-\sum_{d=1}^4 C_{ad}\, 
\langle 0 | \mathrm{T} \psi_{jb}(y) \bar\psi_{id}(x) | 0 \rangle.
\end{equation}
Therefore, the Majorana condition on the propagator matrix is
\begin{equation}\label{majorana-x}
S'(x-y) = C {S'}^T(y-x) C^{-1}.
\end{equation}
Note that the transposition refers to both Dirac and family indices.
In momentum space, equation~(\ref{majorana-x}) reads~\cite{aoki,lavoura}
\begin{equation}\label{majorana}
S'(p) = C {S'}^T(-p) C^{-1}.
\end{equation}
There is the analogous condition for the inverse propagator.

\end{document}